\documentclass[prx,aps,twocolumn,superscriptaddress]{revtex4}

\usepackage{dcolumn}
\usepackage{bm}
\usepackage{amssymb}
\usepackage{color}
\usepackage[pdftex]{graphicx}

\usepackage{amsmath}
\usepackage[linesnumbered,ruled]{algorithm2e}

\begin{document}


\title{Topological Insulator in an Atomic Liquid}

\author{Gia-Wei Chern}
\affiliation{Department of Physics, University of Virginia, Charlottesville, VA 22904, USA}

\date{\today}

\begin{abstract}
We demonstrate theoretically an atomic liquid phase that supports topologically nontrivial electronic structure. A minimum two-orbital model of liquid topological insulator in two dimensions is constructed within the framework of tight-binding molecular dynamics. As temperature approaches zero, our simulations show that the atoms crystallize into a triangular lattice with nontrivial band topology at high densities. Thermal fluctuations at finite temperatures melt the lattice, giving rise to a liquid state which inherits the nontrivial topology from the crystalline phase. The electronic structure of the resultant atomic liquid is characterized by a nonzero Bott index. Our work broadens the notion of topological materials, and points to a new systematic approach for searching topological phases in amorphous and liquid systems. 
\end{abstract}

\maketitle

Topological insulators (TI) are a class of free or weakly-interacting fermionic systems that exhibit topologically nontrivial electron bands~\cite{hasan10,qi11}. One intriguing manifestation of the nontrivial band topology is the appearance of gapless electronic states confined to the sample surface. The study of TI started with the integer quantum Hall (QH) effect, where the Hall conductance is quantized in integer multiples of the universal conductance~\cite{klitzing80}. It was later shown that this quantization is related to a topological index called the TKNN or Chern number of the occupied bands~\cite{thouless82,avron83,haldane88}. Recently, renewed interest in TI was triggered by the discovery of time-reversal invariant systems characterized by a $Z_2$ index~\cite{kane05a}. These include the two-dimensional (2D) quantum spin-Hall insulators~\cite{kane05b,bernevig06a,bernevig06b,konig07} and 3D topological insulators~\cite{fu07,moore07,roy09,hsieh08}. Conventionally, the definition of topological indexes assumes a well-defined Brillouin zone (BZ) from a perfect lattice. For example, the Chern number of TI can be understood as a generalized winding number characterizing the mapping from the BZ to the Hilbert space of Bloch states~\cite{avron83}. Similarly, definitions of the $Z_2$ invariant mostly rely on the condition that  Bloch momentum is a good quantum number~\cite{fu07,moore07,roy09}. Such approaches have enabled a complete classification of TI~\cite{schnyder08,kitaev08,qi08,ryu10} based on the ten symmetry classes of Hamiltonians introduced in the seminal work of Altland and Zirnbauer~\cite{altland97}.

Robustness against disorder is another defining feature of TI.  Effects of disorder on QH insulator as well as TI have been extensively investigated~\cite{sheng97,onoda07,mong12,ringel12,schubert12,kobayashi13,liu16}. Generally speaking, topological properties, such as quantized conductance and gapless boundary modes, are robust against disorder that preserves certain discrete symmetries of the symmetry class, and importantly, provided that the spectral or mobility gap remains finite. Significant efforts have also been devoted to understand disorder-induced non-Anderson-type transitions~\cite{goswami11,pixley15,chen15,syzranov17,kobayashi14,ryu12,shapourian16}. The nontrivial interplay between disorder and band topology also leads to intriguing phenomena such as topological Anderson insulator, in which a metal or an ordinary insulator is transformed into a TI by disorder~\cite{li09,gu10}. Although in most studies, disorder is considered as a perturbation to the otherwise perfectly periodic structure, recent works have demonstrated canonical topological properties in quasi-periodic~\cite{kraus12,fulga16,bandres16} or even amorphous systems~\cite{chern14,agarwala17,mitchell18,poyhonen18}. In light of these findings, complete classification of disordered TI has recently been worked out based on the noncommutative index theorem~\cite{prodan16,katsura18}. These interesting results have significantly broadened the scope of TI and point to new directions in the search of novel topological materials. 

In this paper, we further generalize the notion of topological materials by demonstrating a liquid state of Chern insulator within the framework of tight-binding molecular dynamics (TBMD)~\cite{khan89,wang89,goedecker94}. This novel liquid phase is characterized by a finite spectral gap and a nontrivial Bott index. We consider a 2D system of atoms with active $s$ and $p_x + i p_y$ orbitals. Inter-atomic potentials are parametrized based on the Slater-Koster scheme. At zero temperature, we show that the atoms crystalize into a triangular lattice whose band topology depends on the density. In particular, the lattice is characterized by a quantized non-zero Chern number at high densities. While the triangular lattice is melt by thermal fluctuations at finite temperature in 2D, its nontrivial topology is transferred to the resultant liquid state which is characterized by a  Bott index with the same value as the Chern number in the crystalline phase.

The tight-binding (TB) method has been used to construct several important lattice models for TI, such as the famous Haldane~\cite{haldane88} and Kane-Mele model~\cite{kane05a} on honeycomb lattice, and the Bernevig-Hughes-Zhang (BHZ) model for HgTe quantum well~\cite{bernevig06b}. TB is also the main toolkit for constructing electronic models for disordered, quasi-periodic, and amorphous TIs. In all these systems, a TB Hamiltonian with appropriate parametrization is defined on a fixed network of sites representing the static atomic configuration. On the other hand, the atomic positions are dynamical variables in a liquid and their evolution is usually described using molecular dynamics (MD) simulation. In order to efficiently model the electronic structure of atomic liquids during the dynamical evolution of atoms, here we use the TBMD formulation~\cite{khan89,wang89,goedecker94} to construct a minimal model for liquid Chern insulator. TBMD is one of the so-called quantum MD techniques~\cite{car85,marx09}, in which the atomic forces are obtained by solving the electron wavefunction quantum mechanically at every time-step. Importantly, the electronic structure effects are incorporated into the atomic dynamics in quantum MD methods. 

We consider the following TBMD Hamiltonian, 
\begin{eqnarray}
	\label{eq:H0}
	\mathcal{H} &=& \sum_{ij} \sum_{\alpha\beta} t_{\alpha\beta}(\mathbf r_{ij}) \, c^\dagger_{i\alpha} c^{\,}_{j\beta} 
	+ \sum_i \sum_\alpha \varepsilon_\alpha c^\dagger_{i\alpha} c^{\,}_{i\alpha} \nonumber \\
	& &\qquad + \frac{1}{2}\sum_{i\neq j} \phi(|\mathbf r_{ij}|)
	+\sum_i \frac{|\mathbf p_i|^2}{2m}.
\end{eqnarray}
Here $c^\dagger_{i\alpha}$ is the creation operator of spin-polarized fermions of orbital $\alpha$ at the $i$-th atom, $t_{\alpha\beta}(\mathbf r_{ij})$ is the hopping coefficient, $\phi(r)$ is a classical pair potential for the ions, $m$ is the mass of the ions, and $\mathbf r_{ij} = \mathbf r_j - \mathbf r_i$. The atomic positions $\mathbf r_i$ and momenta $\mathbf p_i$ are treated as classical degrees of freedom.  The first two terms describe an effective TB model for a given atomic configuration $\{\mathbf r_i\}$ and represents the cohesive energy of the system. The third term corresponds to a repulsive energy that originates from the repulsion caused by overlapping orbitals. The last term is the kinetic energy of the atoms. The dynamics of the atoms is described by Newton equation of motion with forces computed quantum mechanically through the Hellman-Feynman theorem: $\mathbf f_i = -\sum_j \rho_{j\beta, i\alpha}\, \partial t_{\alpha\beta}(\mathbf r_{ij})/\partial \mathbf r_i - \sum_j \partial \phi(|\mathbf r_{ij}|)/\partial \mathbf r_i$, where $\rho_{j\beta,i\alpha} = \langle c^\dagger_{i\alpha}c^{\;}_{j\beta} \rangle$ is the electron density matrix solved from the TB model.

Next we describe a minimal 2D atomic system that exhibits nontrivial electronic topological properties. We assume spinless electrons occupying an $s$-like orbital (labeled by $\alpha = 1$) and a $p_x + i p_y$ orbital ($\alpha = 2$). The on-site orbital energies are denoted as $\varepsilon_{1, 2} = \pm \Delta / 2$, where $\Delta$ is the energy splitting of the two orbitals. The hopping matrix can be parameterized using the Slater-Koster formula~\cite{slater54},
\begin{eqnarray}
	t_{\alpha\beta}(\mathbf r) = \left[\begin{array}{cc} t_{ss}(r) & t_{sp}(r)\, e^{i \theta} \\
	-t_{sp}(r)\, e^{-i \theta} & t_{pp}(r) \end{array} \right],
\end{eqnarray}
where $\mathbf r = (x, y)$ is the displacement vector between atoms, $r = |\mathbf r|$, and $\theta = \arctan(y/x)$ indicates the orientation. $t_{ss}$, $t_{sp}$, and $t_{pp}$ correspond to the $(ss\sigma)$, $(sp\sigma)$, and $(pp\sigma)$ bond integrals~\cite{slater54}, respectively.  It is worth noting that same atomic orbitals are used in the $k \cdot p$ theory for the HgTe quantum well that exhibits quantum spin Hall effect~\cite{bernevig06b}.

In applications to real materials, the bond-length $r$ dependence of the hopping $t_{\alpha\beta}(r)$ and pair potential $\phi(r)$ is conventionally determined by fitting {\em ab initio} calculation or experimental data to specific functional forms~\cite{goodwin89}. As a toy model to demonstrate the proof of principle, here we assume simple exponential decay for both functions: $t(r) = V \,\exp(-r/ d)$ and $\phi(r) = \phi_0 \,\exp(-r/ \xi)$. More specifically, we assume the same characteristic length $d$ for the three $(ss\sigma)$, $(sp\sigma)$, and $(pp\sigma)$ bonds. Specifically, we used $\xi = 0.54 d$, and $\phi_0 = 33.4\,\Delta$, $V_{ss} = -V_{pp} = V_{sp} = 4.2 \,\Delta$. In the following, we use $d$ and $\Delta$ for units of length and energy, respectively. We have checked that the existence of the liquid topological state does not depend sensitively on these parameters. However, the condition $V_{ss} \, V_{pp} < 0$, which is satisfied in most materials, is crucial to the realization of topological phase. Finally, we assume half-filling for all simulations.

\begin{figure}
\includegraphics[width=0.99\columnwidth]{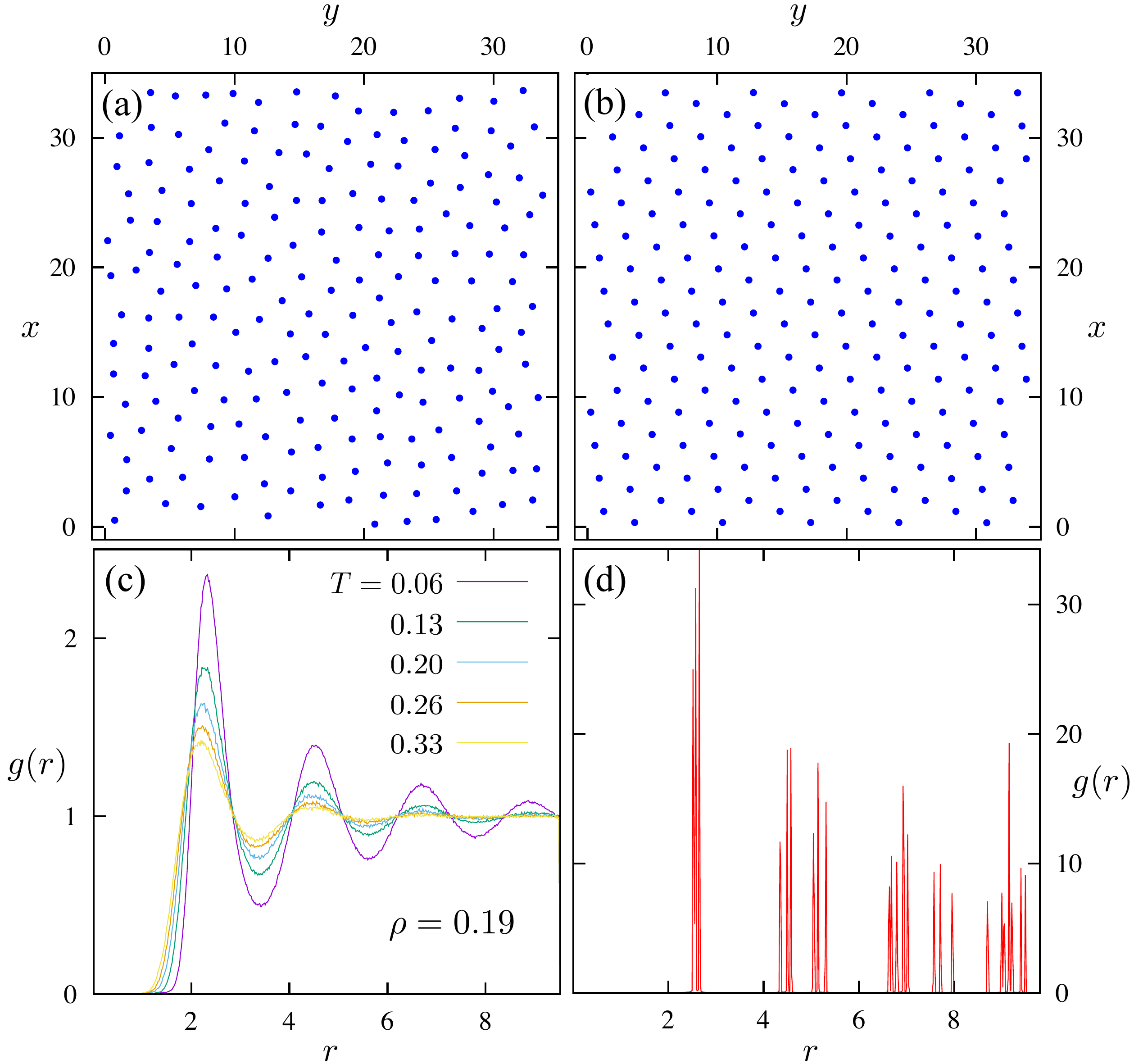}
\caption{(Color online)  
\label{fig:configs} TBMD simulations of model Hamiltonian Eq.~(\ref{eq:H0}) with $N = 200$ atoms and a fixed particle density $\rho = 0.19$. Snapshots of atomic positions at temperatures $T = 0.06$ and $T = 10^{-5}$ are shown in (a) and (b), respectively. The thermal-averaged pair distribution functions $g(r)$ for the liquid and the (quasi) crystalline phases are shown in panels~(c) and (d), respectively.
}
\end{figure}

We performed our TBMD simulations using the standard velocity Verlet algorithm~\cite{allen87}. The interatomic forces are computed by exactly diagonalizing the TB Hamiltonian at every time step. Fig.~\ref{fig:configs}(a) and (b) show sample snapshots at temperatures $T = 0.06$ and $T = 10^{-5}$, respectively.  The atoms at high temperatures are disordered. Yet they exhibit strong short-range correlation as evidenced from the radial distribution function $g(r)$ at finite temperatures shown in Fig.~\ref{fig:configs}(c). Both the vanishing $g(r)$ for $r \lesssim 1.8$ and the oscillatory feature are characteristic of a liquid phase. As $T$ goes to zero, the atoms crystallize into a triangular lattice as shown in Fig.~\ref{fig:configs}(b), which also manifests itself in the delta-like peaks of the corresponding $g(r)$ [Fig.~\ref{fig:configs}(d)]. Strictly speaking, thermal fluctuations preclude any crystalline order in 2D. Indeed, the delta-like peaks disappear even at temperatures as low as $T \sim 10^{-3}\Delta$. However, it is appropriate to consider local crystalline order at low enough temperatures when the thermal correlation length $\xi_T$ is much larger than the system size.

\begin{figure}
\includegraphics[width=0.99\columnwidth]{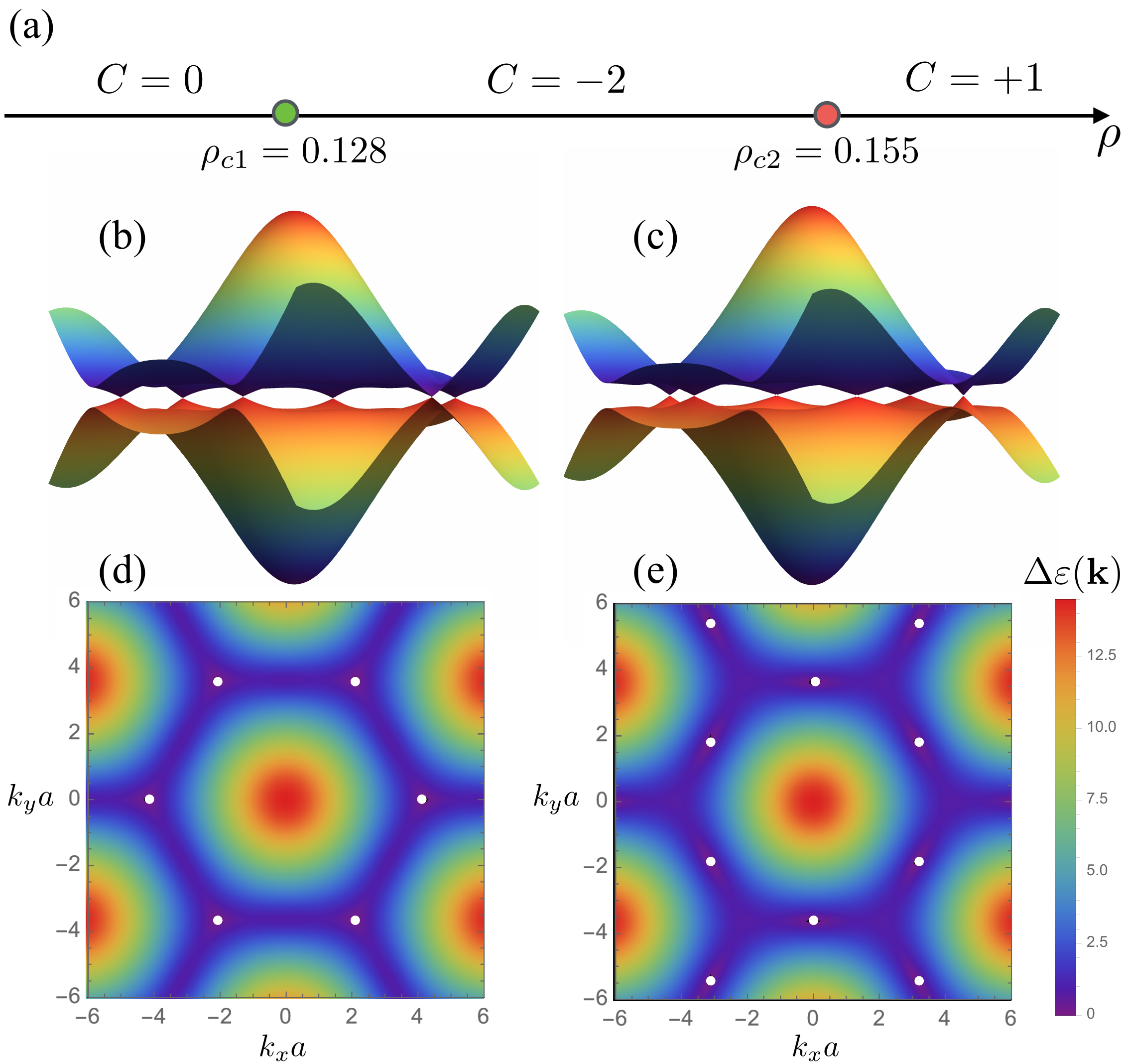}
\caption{(Color online)  
\label{fig:bands} (a) phase diagram showing two quantum critical points separating three topologically distinct phases in the triangular-lattice TB model. Panels (b) and (c) show the band structures at the two critical densities $\rho_{c1}$ and $\rho_{c2}$, respectively. Six Dirac nodes develop at (b) the corners (two inequivalent $K$ points) of the BZ, and (c) the midpoints of BZ edges (three inequivalent $M$ points). The $\mathbf k$-dependent band-gap $\Delta\varepsilon(\mathbf k) = 2 |\mathbf m(\mathbf k)|$ at the two critical densities are shown in (d) and (e), respectively. The white dots mark the Dirac points where $\Delta\varepsilon =0$.
}
\end{figure}

The crystallization of atoms at $T \to 0$ makes it easier to analyze their topological properties using conventional band theory toolkits. First, the lattice TB model can be easily solved for rather large lattices using the Fourier transform method. The TB Hamiltonian in momentum space reads $H_{\alpha\beta}(\mathbf k) = \sum_{\mathbf R} t_{\alpha\beta}(\mathbf r)\,\exp(i \mathbf k\cdot\mathbf R)$, where $\mathbf R$ runs over the triangular lattice sites up to a large cutoff $r_c = 20$. To obtain its spectrum and topological index, we express the $2\times 2$ Hamiltonian in terms of the Pauli matrices: $H(\mathbf k) = \varepsilon_0(\mathbf k) \mathbb I + \mathbf m(\mathbf k) \cdot \bm \sigma$, where $\mathbb I$ is a $2\times 2$ identity matrix. Its spectrum consists of two bands with dispersions readily computed as: $\varepsilon_{\pm}(\mathbf k) = \varepsilon_0(\mathbf k) \pm |\mathbf m(\mathbf k)|$. In most cases, the two bands do not overlap and are separated by a direct energy gap $\varepsilon_g = 2 \min_{\mathbf k} |\mathbf m(\mathbf k)|$. At half-filling, the lower energy band $\varepsilon_{-}(\mathbf k)$ is completely filled. Our explicit numerical calculation shows that the band-gap closes at two critical densities, at which low-energy Dirac cones emerge at the high-symmetry points of the Brillouin zone (BZ) as shown in Fig.~\ref{fig:bands}.

The two critical points $\rho_{c1, 2}$ also separate phases of different topologies. In the gapped phase, the Chern number of the occupied band is given by the integral over the BZ: $C = (1/4\pi) \int\int \hat{\mathbf m}\cdot \partial_{k_x} \hat{\mathbf m} \times \partial_{k_y} \hat{\mathbf m} \, dk_x dk_y$. At low densities $\rho < \rho_{c1}$, we find a trivial band insulator with $C = 0$. The band-gap closes at the first critical density $\rho_{c1}$ as Dirac cones develop at the six corners of the BZ. Only two of these six Dirac nodes ($K$ and $K'$) are inequivalent. Further increasing the density gaps out these two Dirac nodes, giving rise to a topological insulator with $C = -2$. As we further increase the density, the atoms undergo another quantum phase transition at $\rho_{c2}$ characterized again by appearance of Dirac cones, this time at the midpoints of BZ edges; these correspond to three inequivalent $M$ points. Gapping out these three Dirac nodes increases the Chern number by 3, leading to another topological insulating phase with $C = +1$.

Having described the topological properties of the crystalline phase in the $T \to 0$ limit, one natural question that arises is how the nontrivial band topology is transferred to the liquid state at finite temperatures. To answer this question, we first compute the average energy gap of the atomic fluid. For a system with $N$ atoms, solving the TB Hamiltonian in Eq.~(\ref{eq:H0}) gives $2N$ eigen-energies $\varepsilon_m$ ($m = 1, 2, \cdots, 2N$). Assuming these energies are in ascending order, we define an energy gap at half-filling as $\varepsilon_g = \varepsilon_{N+1} - \varepsilon_N$. We note that at finite $T$, there are always thermal excitations of electrons above the gap, so this $\varepsilon_g$ should be viewed as the intrinsic energy gap of the electron spectrum, in the same spirit as the definition of band-gap for a crystalline state. 
A crucial difference is that a configurational average is required in order to obtain an effective energy gap for the liquid phase.

\begin{figure}
\includegraphics[width=0.99\columnwidth]{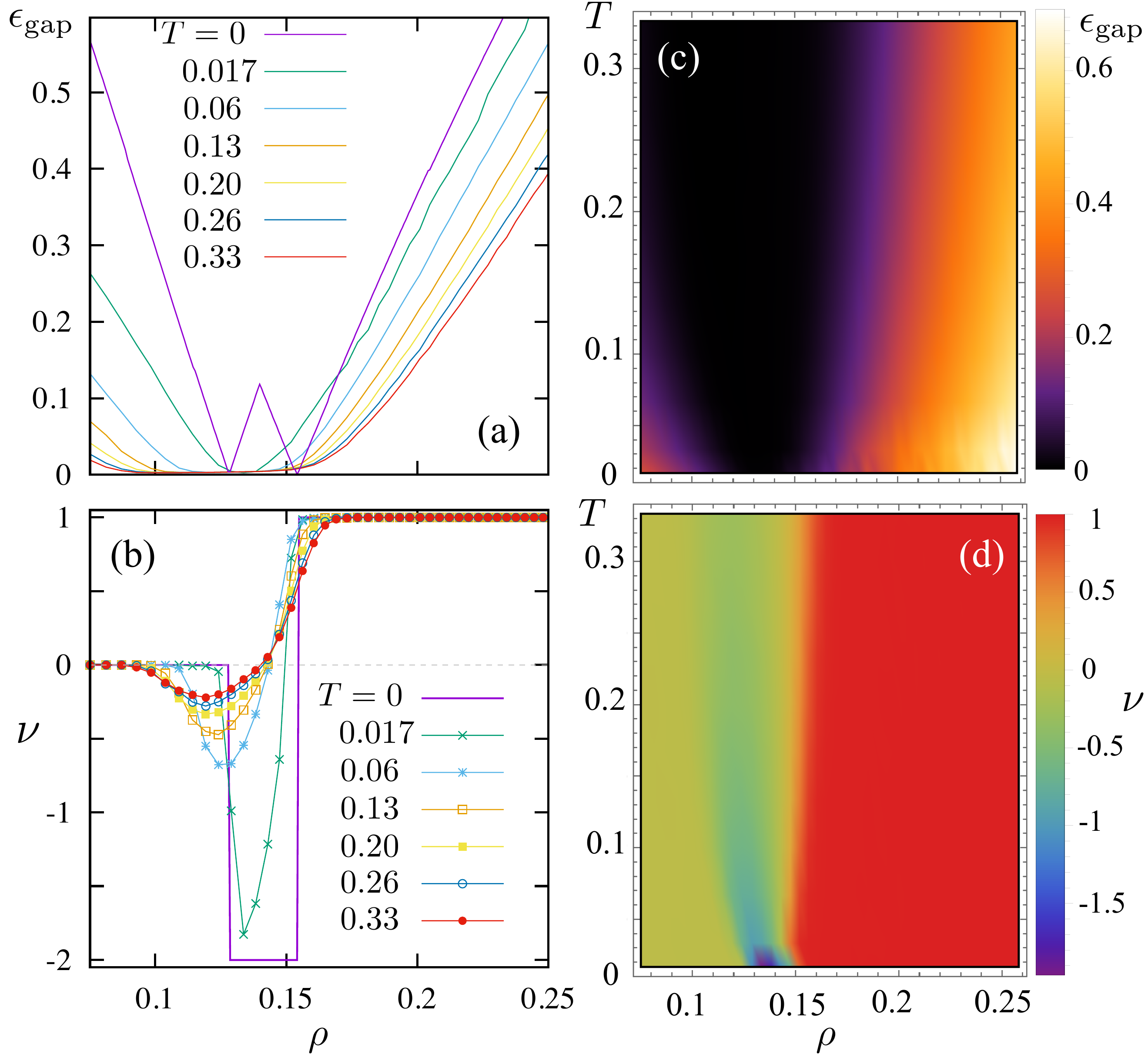}
\caption{(Color online)  
\label{fig:phase-diagram} (a) Energy gap $\varepsilon_g$ as a function of density for varying temperatures in the liquid phase. Also shown for comparison is the band-gap of the crystalline phase obtained using Fourier transform method. (b)~Configuration-averaged Bott index $\nu$ vs density $\rho$ for varying temperatures. The $T = 0$ curve is the Chern number $C$ of the triangular-lattice TB model. Panels (c) and (d) show the density plot of averaged energy gap and Bott index in the $T$-$\rho$ plane.
}
\end{figure}

Fig.~\ref{fig:phase-diagram}(a) shows the density dependence of the energy gap $\varepsilon_g$ averaged over a large number of atomic configurations in the liquid state at different temperatures. Also shown for comparison is the curve for the crystalline phase. As discussed above, $\varepsilon_g$ vanishes at two critical densities $\rho_{c1, 2}$ in the presence of crystalline order. 
In the liquid phase, the energy gap becomes very small for a finite range of density around $\rho_{c1}$, indicating an extended regime of liquid metal phase. Moreover, the size of this metallic regime increases with increasing temperature. On the other hand, a nonzero gap persists at small and large densities even in the liquid state.

To describe the nontrivial topology of the atomic liquid, we compute a topological invariant called the Bott index~\cite{loring10}, which plays a similar role as the Chern number for crystalline state. Indeed, the Bott index reduces to Chern number in the presence of long-range crystalline order. To compute this index for a given atomic configuration $\{\mathbf r_i = (x_i, y_i)\}$, we first construct two diagonal matrices $\Theta$ and $\Phi$, such that $\Theta_{ii} = 2\pi x_i / L_x$ and $\Phi_{ii} = 2\pi y_i /L_y$. Next we compute the band-projected position matrices $U$ and $V$ according to $U = P \exp(i \Theta) P$ and $V = P \exp(i \Phi) P$, where $P$ is the projector to the lowest $N$ eigen-states, corresponding to the (half) occupied states at zero temperature. The Bott index is then given by $\nu = (1/2\pi)\, {\rm Im}\{{\rm tr}[\log(VUV^\dagger U^\dagger)]\}$ which only takes on integer values~\cite{loring10}. Essentially, $\nu$ quantifies the topological obstructions to the existence of localized Wannier functions spanning the occupied manifold~\cite{loring10}.

\begin{figure}[t]
\includegraphics[width=0.99\columnwidth]{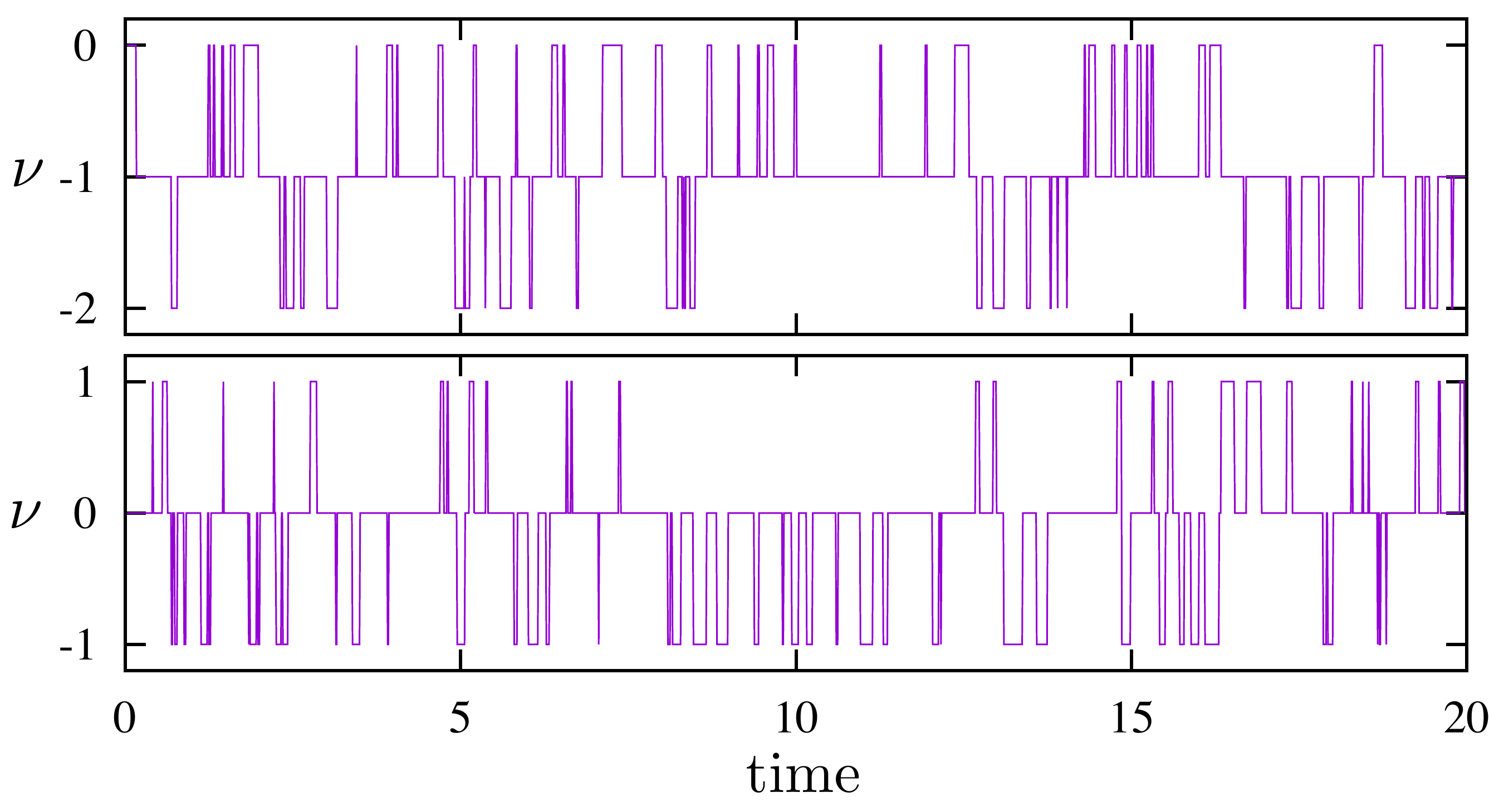}
\caption{(Color online)  
\label{fig:bott-time} The instantaneous Bott index as a function of time for (a) $\rho = 0.133$ and (b) $\rho = 0.143$ from TBMD simulations with temperature $T = 0.017$.
}
\end{figure}

By averaging over a large number of atomic configurations from our TBMD simulations, we compute the Bott index as a function of density for various temperatures in the liquid phase. The results are summarized in Figs.~\ref{fig:phase-diagram}(b) and (d). Also shown for comparison is the numerical Chern number of the crystalline state. At small and large densities, the averaged $\nu$ exhibits quantized values at 0 and 1, respectively. Taking into account that a finite gap exists also in these two regimes, we conclude that the atomic liquid is a trivial band insulator at small density, and a topological Chern insulator at large~$\rho$. In fact, we find that $\nu$ is exactly quantized at 0 and 1 for small and large $\rho$, respectively, for all snapshots in our MD simulations. 
For intermediate values of $\rho$, the Bott index fluctuates between different integer values, as illustrated in Fig.~\ref{fig:bott-time}, giving rise to a non-quantized $\nu$ after average. This intermediate phase partially inherits the nontrivial topology of the crystalline state with a Chern number $C = -2$ and can be viewed as a topological liquid quasi-metal. Its electronic structure might be similar to the diffusive metal phase induced by strong disorder in topological semimetals~\cite{goswami11,pixley15,chen15,syzranov17,kobayashi14,ryu12,shapourian16}.

Our two-orbital TBMD model provides the first example of a Chern insulator or quantum anomalous Hall insulator in an atomic liquid. One can use this model to construct a liquid state of quantum spin-Hall (QSH) insulator similar to how Kane-Mele model is built from the Haldane model~\cite{kane05a}.  Since time-reversal symmetry is required for the QSH effect, we add spins to the electrons and consider two copies of our atomic model: one with $|s, \uparrow\rangle$ and $|p_x + i p_y, \uparrow \rangle$ spin-orbital states, while the other with $|s, \downarrow\rangle$ and $|p_x + i p_y, \downarrow \rangle$. Again, the same atomic orbitals are used in the BHZ model for the quantum spin-Hall insulator in HgTe quantum wells~\cite{bernevig06b}. This simple model with two decoupled copies related by time-reversal symmetry gives a quantized spin-Hall conductivity. In the presence of small spin-nonconserving perturbations, the topological nature of the quantum spin-Hall state ensures the existence of spin-filtered edge states  as long as time-reversal symmetry is preserved. Construction of 3D liquid TI will be left for future studies.

To summarize, we have demonstrated a novel liquid topological insulator based on the TBMD formulation. This new state of matter generalizes the notion of topological materials, and offers intriguing technological possibilities. Together with the recent reports of TI in amorphous systems~\cite{agarwala17,mitchell18,poyhonen18}, our work further shows the robust nature of topological insulators, which does not rely on the existence of a well defined BZ. Indeed, the amorphous TI phase can be viewed as the high-$T$ ``gas" phase, characterized by a flat radial distribution function, of our atomic system. Given the fact that TI can exist even in a set of randomly connected points~\cite{agarwala17}, the possibility of a liquid TI might not be a complete surprise. However, it is unclear whether the topological properties survive once these ``points" or atoms are allowed to move around. For example, the rearrangement of atoms might close the electronic gap, rendering the system a liquid metal. Our proof of principle study shows that nontrivial topology of electronic structure can exist in a dynamical liquid state.

Our work also suggests that systematic searches for liquid or amorphous TIs can be attained by the TBMD approach. Over the past few decades, significant progress has been made toward TBMD modeling of realistic materials such as transition metals, semiconducting alloys, and liquid chalcogenides~\cite{cleri93,goringe97}, especially with more accurate parameterization of the radial hopping functions and pair potentials~\cite{goodwin89}. However, the effects of spin-orbit coupling, which is crucial to topological properties, have yet to be systematically investigated in the TBMD framework. An important step towards this goal is the TBMD parametrization of systems containing $4d$ or $5d$ atoms, which play an key role in most topological materials. Interestingly, several well-studied TI compouds, such as Bi$_2$Te$_3$, Sb$_2$Te$_3$, Ag$_2$Te, Ag$_2$Se, and PbTe, are listed as liquid semiconductors in their molten state~\cite{enderby90}. Such liquid alloys exhibit semiconducting behaviors and are characterized by short-range order and a well defined chemical stoichiometry. It is intriguing to see whether these compounds are potential liquid TI through careful {\em ab initio} MD simulations~\cite{marx09} and characterization. We hope that our work will motivate further numerical and experimental search of such liquid TI in real materials. 


\bigskip

{\em Acknowledgements}. I thank Jeffrey Teo, D.~Louca, and U.~Chatterjee for insightful discussions on disordered topological insulators and collaborations on related projects. This work is partially supported by the Center for Materials Theory as a part of the Computational Materials Science (CMS) program, funded by the US Department of Energy, Office of Science, Basic Energy Sciences, Materials Sciences and Engineering Division.

\end{document}